\documentclass[twocolumn,english,prb,showpacs]{revtex4}
\usepackage[latin9]{inputenc}
\usepackage{amssymb}
\usepackage{graphicx}
\usepackage{amsmath}
\usepackage{color}
\usepackage{mathrsfs}
\usepackage{float}
\usepackage{indentfirst}
\usepackage{mathrsfs}	
\usepackage{float}
\usepackage{indentfirst}
\usepackage{textcomp}

\newcommand\ccite[1]    {~\cite{#1}}
\newcommand\PLOTFILE[1]  {./#1}

\newcommand{\vx}{{\mathbf x}}

\newcommand{\COMMENTED}[1]{}

\begin{document}
\title{Symmetry-projected Wave Functions in Quantum Monte Carlo
  Calculations}

\author{Hao Shi$^{1}$, Carlos A. Jim\'enez-Hoyos$^{2}$,
  R. Rodr\'{\i}guez-Guzm\'an$^{2,3}$, Gustavo E. Scuseria$^{2,3}$ and
  Shiwei Zhang$^{1}$}

\affiliation{$^{1}$ Department of Physics, The College of William and
  Mary, Williamsburg, Virginia 23187 \\ $^{2}$ Department of
  Chemistry, Rice University, Houston, Texas 77005, USA \\ $^{3}$
  Department of Physics and Astronomy, Rice University, Houston, Texas
  77005, USA}

\begin{abstract}
We consider symmetry-projected Hartree--Fock trial wave functions in
constrained-path Monte Carlo (CPMC) calculations. Previous CPMC
calculations have mostly employed Hartree--Fock (HF) trial
wave functions, restricted or unrestricted.  The symmetry-projected HF
approach results in a hierarchy of wave functions with increasing
quality: the more symmetries that are broken and restored in a
self-consistent manner, the higher the quality of the trial
wave function.  This hierarchy is approximately maintained in CPMC
calculations: the accuracy in the energy increases and the statistical
variance decreases when further symmetries are broken and
restored. Significant improvement is achieved in CPMC with the best
symmetry-projected trial wave functions over those from simple HF.  We
analyze and quantify the behavior using the two-dimensional repulsive
Hubbard model as an example.  In the sign-problem-free region, where
CPMC can be made exact but a constraint is deliberately imposed here,
spin-projected wave functions remove the constraint bias.  Away from
half-filling, spatial symmetry restoration in addition to that of the
spin lead to highly accurate results from CPMC.  Since the
computational cost of symmetry-projected HF trial wave functions in
CPMC can be made to scale algebraically with system size, this
provides a potentially general approach for accurate calculations in
many-fermion systems.
\end{abstract}


\maketitle

\section{Introduction}\label{sec:Introduction}

Approximations able to account for the most relevant correlations in
low-dimensional many-fermion systems represent a cornerstone in
condensed matter physics.  In particular, very challenging phenomena
such as the high-Tc superconductivity,\ccite{HTC_Bednorz} the colossal
magnetic resistance\ccite{Science-Dagotto} as well as the
superconductivity in the iron-based
compounds,\ccite{sup-Fe,Stewart-Review} just to mention three of the
most notable examples, require us to better understand the nature of
the electron-electron correlations and their impact on the resulting
properties of the considered quantum systems. Within this context, the
Hubbard model is regarded as a paradigm since, in spite of its
essential simplicity, it displays several challenging properties
associated with the relevant physics in strongly correlated
many-electron systems. From the theoretical point of view,
Hubbard-like models also represent valuable benchmark systems for
testing different state-of-the-art approximations that can
subsequently be extended to more realistic situations.

Nowadays there are several such approximations already used to study
both the one (1D) and two-dimensional (2D) Hubbard models with
variable degrees of success. For sufficiently small lattices, exact
diagonalization is possible. However, due to its exponential cost,
such an exact diagonalization becomes impossible beyond a given system
size. One can then resort to other approximations such as the
variational Monte
Carlo\ccite{VMCPhysRev.138.A442,VMCPhysRevB.16.3081,VMCPhysRevLett.60.1719}
(VMC), the Coupled Cluster\ccite{coupledcluster} (CC) and the Density
Matrix Renormalization
Group\ccite{DMRGPhysRevLett.69.2863,RevModPhys.77.259,Schollwock201196}
(DMRG) methods.

Auxiliary-field quantum Monte
Carlo\ccite{BSS1PhysRevD.24.2278,KooninSugiyama19861,SZ_2013LectNotes}
(AFQMC) is one of the most popular methods to extract collective
properties of quantum many-body systems. However, it is limited by a
sign problem in Hubbard-like models, where the local interactions lead
to auxiliary-fields that are real, and by a phase problem in realistic
electronic systems, where the Coloumb interaction leads to complex
fields.\ccite{SZ_2013LectNotes} In order to remove the exponential
increase of the statistical noise with system size (or inverse
temperature), we control the sign/phase problem approximately.  The
quantum Monte Carlo (QMC) process is formulated as open-ended random
walks in the space of Slater determinants. A constraint, formally
exact, is imposed on the paths of the random walk according to a sign
or phase condition of the walker.  This approach has been referred to
as the constrained path Monte
Carlo\ccite{PhysRevB.55.7464,PhysRevB.78.165101} (CPMC) method in the
case of ``only'' a sign problem and the phaseless
AFQMC\ccite{PhysRevLett.90.136401} for the general case with a phase
problem.  In practice, the constraint is implemented by the overlap of
the sampled Slater determinants with a trial wave function. Despite its
approximate nature, this approach is very accurate even with a simple
mean-field trial wave function as it has been shown before in a wide
range of
applications.\ccite{PhysRevB.80.214116,PhysRevLett.104.116402,WirawanJCP2011,SZ_2013LectNotes}

In quantum mechanical systems, the symmetries of the Hamiltonian are
crucial in characterizing properties such as the excitation spectrum,
the basic fingerprint of the
system.\ccite{nuclearsym,ImadaPRB2004,PhysRevB.72.224518,PhysRevB.72.085116,PhysRevE.77.026705,JPSJ.77.114701,PhysRevB.85.245130}
We have studied\ccite{PhysRevB.88.125132} how the use of symmetry
influences the CPMC method and AFQMC calculations in general, both in
the form of the Hubbard-Stratonovich (HS) transformation and in the
trial wave function. It was shown that imposition of symmetry
properties resulted in a significant increase in the accuracy and
efficiency in the QMC.  Choosing a symmetry-adapted HS transformation
often does not change the cost of the calculation significantly. On
the other hand, symmetry-adapting a trial wave function may result in
expansions whose length depend on system size. In our prior work, we
obtained symmetry-adapted wave functions by using a complete active
space (CAS) expansion in the subspace of the open shell. The problem
with CAS expansions is that the size of the wave function scales
exponentially with the dimension of the active space.

An alternative to obtain highly correlated wave functions which, at the
same time, respect the original symmetries of the considered
many-fermion problem is provided by symmetry restoration via
projection techniques. Very recently, a hierarchy of variational
symmetry-projected approximations, inherited from nuclear structure
physics,\ccite{nuclearsym} has been considered to describe the ground
and excited states of the 1D (Ref. \onlinecite{PhysRevB.87.235129})
and 2D (Ref. \onlinecite{PhysRevB.85.245130}) Hubbard models. They
have also been successfully incorporated into the description of
molecular systems within the Projected Quasiparticle Theory (PQT)
framework.\ccite{JCP.135.124108,JCP.136.164109} A symmetry-projected
HF wave function considers a Slater determinant $|\bar \phi \rangle$
which deliberately breaks several symmetries of the original
Hamiltonian. By the superposition of the (degenerate) Goldstone
manifold $\hat{R} | \bar \phi \rangle$, where $\hat{R}$ represents a
symmetry operation, one can recover the desired symmetry in a fully
self-consistent variation-after-projection (VAP) procedure. In
addition, the intrinsic Slater determinant $|\bar \phi \rangle$
resulting from the symmetry-projected VAP procedure contains defects
that provide an intuitive physical picture regarding the basic units
of quantum fluctuations in the considered
systems.\ccite{PhysRevB.87.235129} Recently symmetry-restored trial
wave functions have been used in CPMC in the study of nuclear shell
models.\ccite{CPMC_sym_shell_PRL13}

It is known that symmetry-projection out of a Slater determinant
results in a wave function that is not size
extensive,\ccite{JCP.136.164109} {\em i.e.}, the correlation energy
recovered by such an ansatz does not scale linearly with system
size. This is also true for the CAS expansions we have considered
before. Symmetry-projected wave functions can, nevertheless, account
for some of the strong correlations (due to the degeneracies in the
single-particle spectrum).  In this paper, we examine the use of
symmetry-projected wave functions as trial states within the CPMC
framework.  We stress that, even though a single symmetry-broken
Slater determinant is used in the trial wave function, the
symmetry-projected wave functions are, via the projection operators,
multi-determinant in nature. In addition, the symmetry-projected
state can access highly-excited configurations. In this way, they
bring a very sophisticated trial ansatz which can be expected to
improve the quality of the sign/phase constraint in CPMC.  Indeed, our
results show that the use of such trial wave functions greatly improves
the CPMC results. The combined approach demonstrates dramatically
better behavior that successfully removes the size-extensivity problem
of projected HF. Often the full symmetry-projected trial wave function
eliminates the systematic error from the constraint, and the CPMC
calculations approach the exact answer, with a reduced variance.

\section{Method}

In this work, we use as benchmark system the two-dimensional repulsive
Hubbard model, which is written in second-quantized form
as\ccite{J.hubbard.1963}
\begin{equation}
 \hat{H}=\hat K +\hat V=-t\sum_{\langle {\bf{i}}, {\bf{j}} \rangle
   \sigma}^{L} c_{{\bf{i}} \sigma}^{\dagger}c_{{\bf{j}} \sigma}
 +U\sum_{{\bf{i}}}^{L} n_{{\bf{i}}\uparrow}n_{{\bf{i}}\downarrow}\,,
 \label{eq:HubHamil}
\end{equation}
where the first term represents kinetic energy from electron hopping
($t > 0$) and the second is the repulsive on-site interaction ($U >
0$).  The operators $c_{{\bf{i}}\sigma}^{\dagger}$ and
$c_{{\bf{i}}\sigma}$ create and annihilate an electron with spin
direction $\sigma$ on the ${\bf{i}}$-th lattice site. Note that we
have introduced a vector notation for the site index, {\em i.e.},
${\bf{i}}=(i_{x},i_{y})$.  The operators $\hat{n}_{{\bf{i}} \sigma}$ =
$\hat{c}_{{\bf{i}} \sigma}^{\dagger} \hat{c}_{{\bf{i}} \sigma}$ are
the local number operators. The notation $\langle {\bf{i}}, {\bf{j}}
\rangle$ is used to denote that only hopping between nearest-neighbor
sites is allowed. We assume periodic boundary conditions (PBC) in both
the x and y directions. Throughout this paper, energies will be
reported in units of $t$ and we set $t=1$.

\subsection{Symmetry-Projected Wave function}

In this section, we briefly discuss the form of the symmetry-projected
HF states used as trial wave functions. For more details of the
symmetry-projected approach, we refer the reader to our previous works
(Refs. \onlinecite{PhysRevB.87.235129} and
\onlinecite{PhysRevB.85.245130}).

We work with HF-type Slater determinants which preserve the number of
electrons in the system. An arbitrary HF-type Slater determinant can
in general break several symmetries of the original Hamiltonian.
Typical examples are the rotational (in spin space) and spatial (space
group for periodic systems) symmetries. In the present study, we have
considered two different kinds of symmetry broken states:
(unrestricted) UHF configurations which are eigenstates of $S_{z}$ and
therefore have a definite $N_\uparrow$ and $N_\downarrow$, as well as
(generalized) GHF configurations that break $S_z$ and therefore can
only be characterized by an overall $N$ number of electrons.

To restore the spin quantum numbers, we use the spin projection
operator
\begin{eqnarray} \label{PROJ-S} 
\hat{P}_{\Sigma {\Sigma}^{'}}^{S} = \frac{2S+1}{8 {\pi}^{2}} \int d
\Omega \, {\cal{D}}_{\Sigma {\Sigma}^{'}}^{S *} (\Omega)
\hat{R}(\Omega),
\end{eqnarray}
where $\hat{R}(\Omega)= e^{-i \alpha \hat{S}_{z}} e^{-i \beta
  \hat{S}_{y}} e^{-i \gamma \hat{S}_{z}}$ is the rotation operator in
spin space, the label $\Omega = \left(\alpha, \beta, \gamma \right)$
stands for the set of Euler angles, and ${\cal{D}}_{\Sigma
  {\Sigma}^{'}}^{S}(\Omega)$ are Wigner functions. We note that, if
UHF wave functions are used, the numerical effort can be alleviated
significantly in the evaluation of matrix elements, as both the
integrals over $\alpha$ and $\gamma$ can be carried out analytically.

To recover the spatial symmetries we use the generic projection
operator
\begin{eqnarray} \label{proj-space-group}
\hat{P}_{m m^{'}}^{k} = \frac{l}{h} \sum_{g} 
{\Gamma}_{m m^{'}}^{k\ast}(g) \hat{R}(g),
\end{eqnarray}
where $\Gamma^{k}$ is an irreducible representation (irrep), which can
be found by standard methods, and $\hat{R}(g)$ represents the
corresponding symmetry operations in the considered lattices
parametrized in terms of the label $g$. In addition, $l$ is the
dimension of the irrep and $h$ is the order of the group in
Eq.(\ref{proj-space-group}). We note that, for the periodic Hubbard 2D
system, the irreducible representations associated with the spatial
symmetry can be characterized by the linear momentum ($k$) inside the
Brillouin zone and, for certain high-symmetry momenta, by additional
parities under $x$, $y$, and $x$-$y$ reflections. In what follows, we
will explicitly provide the linear momentum of the recovered irrep
and, where appropriate, we will further provide the parities of the
recovered state (if the full space group projection is carried out).

We then superpose the Goldstone manifolds due to spin and/or spatial
symmetries via the following ansatz

\begin{eqnarray} \label{SP-PHF}
| \Psi_{K}^{\Theta} \rangle = \sum_{K^{'}} f_{K^{'} }^{ \Theta }
\hat{P}_{K K^{'}}^{\Theta} | \bar \phi \rangle,
\end{eqnarray}
where $f^\Theta_{K^{'}}$ are variational parameters, and $|\bar
\phi\rangle$ is the reference HF-type single Slater determinant. Here,
the indices $\Theta$ and $K$ denote quantum numbers associated with
spin and/or spatial symmetries. Let us stress that, through the action
of the projection operator $\hat{P}_{K K^{'}}^{\Theta}$, the
multideterminantal character of the state characterized by the quantum
numbers $\Theta$ and $K^{'}$ is recovered and written in terms of the
quantum numbers $\Theta$ and $K$.  The linear combination introduced
in Eq.(\ref{SP-PHF}) guarantees the independence of
$|\Psi_{K}^{\Theta} \rangle$ with respect to rotations of the Slater
determinant $|\bar \phi \rangle$.

The wave functions $| \Psi_{K}^{\Theta} \rangle$ of Eq.(\ref{SP-PHF})
are precisely the ones used in the present study as trial states
within the CPMC scheme. The wave functions, Eq.(\ref{SP-PHF}), are
determined by resorting to the Ritz-variation of the projected energy
with respect to both the mixing coefficients ($f^\Theta_{K^{'}}$) and
the HF-type determinant $|\bar \phi \rangle$. We stress that the
wave functions are fully optimized in the presence of the projection
operations, i.e., a VAP approach. In our previous
work,\ccite{PhysRevB.87.235129} we have discussed how the optimized
determinant $|\bar \phi \rangle$ develops defects that can be
interpreted as the basic units of quantum fluctuations in the
system. We refer the reader to Ref. \onlinecite{PhysRevB.85.245130}
for a discussion of how the optimization procedure is carried out in
practice.

Before concluding this section, let us introduce the notation used in
this paper for the symmetry-projected wave functions of
Eq.(\ref{SP-PHF}):

\begin{itemize}

\item LM (linear momentum) is used to denote wave functions in which
  the linear momentum has been broken and restored.

\item SG (space group) is used for wave functions in which the full
  space group of the 2D square lattice has been broken and restored.

\item S$_z$ is used to indicate that $S_z$ projection has been done
  (out of a GHF-type determinant).

\item S (spin) is used to indicate that full spin projection has been
  carried out. If the determinant is of GHF-type (noncollinear), it
  will involve the triaxial integration of Eq.(\ref{PROJ-S}).

\end{itemize}

For example, SG,S-UHF means that the trial wave function was prepared
by breaking and restoring spin and space group from a UHF-type
determinant, while LM,S$_z$-GHF indicates that linear momentum and
$S_z$ have been broken and restored out of a GHF-type determinant. In
both cases, the determinant is variationally optimized to minimize the
energy in the presence of the symmetry projection operators, and
before introducing the Monte Carlo procedure described in the next
section.

\subsection{CPMC}
We briefly summarize the ground-state CPMC
method\ccite{PhysRevB.55.7464,PhysRevB.78.165101} below to facilitate
the ensuing discussions. The reader is referred to
Ref.~\onlinecite{SZ_2013LectNotes} and references therein for further
details. Aside from the symmetry-projected trial wave function, there
is a new algorithmic element in the present study, namely we have
generalized the AFQMC approach to have GHF-type of random walkers, as
discussed in Sec.~\ref{ssec:away-half-filling}.  This will allow the
application of CPMC and AFQMC in general to Hamiltonians that do not
conserve $\hat S_z$, for example one that includes spin-orbit
coupling.

All ground-state AFQMC methods are based on the projection
\begin{equation}
 |\Psi_{0}\rangle \propto \lim_{\beta\rightarrow\infty} e^{-\beta (\hat{H}-E_{T})} |\Psi_{T}\rangle\,,
\end{equation}
where $E_{T}$ is a trial ground-state energy (an initial guess which
is then improved in the calculation) and $|\Psi_{T}\rangle$ is a guess
of the ground state wave function, a trial wave function, which is
typically taken to be a single Slater determinant or a linear
combination of Slater determinants.  $\langle
\Psi_{0}|\Psi_{T}\rangle\neq0$ needs to be satisfied in order to
project to the ground state.  In a numerical method, the limit can be
obtained iteratively by
\begin{equation}
  |\Psi^{(n+1)}\rangle = e^{-\Delta\tau {\hat H}}|\Psi^{(n)}\rangle,
\label{eq:process}
\end{equation}
where $|\Psi^{(0)}\rangle = |\Psi_T\rangle$ and $\Delta\tau$ is a small positive number. 

The Trotter-Suzuki breakup\ccite{trotter1959,MasuoSuzuki1976} and a
so-called Hubbard-Stratonovich (HS) transformation are used to break
the propagator into many one-body propagators\ccite{SZ_2013LectNotes}
\begin{equation}
      e^{-\Delta\tau {\hat H}}= \int d\vx\, p(\vx) {\hat B}(\vx),
\label{eq:HS}
\end{equation}
where $p(\vx)$ is a probability density function, for example, a
uniform distribution of one Ising field per site ($x_{\bf{i}}=\pm1$ with
${\bf{i}}=1,2,\cdots,L$.).  The one-body propagator ${\hat B}(\vx)$ in
Eq.~(\ref{eq:HS}) has a {\em special form\/}, namely, a product of the
exponential of one-body operators
\begin{equation}
\hat B(\vx)= e^{-\Delta\tau {\hat K}/2}\,e^{\hat v(\vx)}e^{-\Delta\tau
  {\hat K}/2}\,
\label{eq:Bx}
\end{equation}
where ${\hat v(\vx)}$ is a one-body operator whose matrix elements are
simple functions of the HS fields and of magnitude ${\mathcal
  O}(\sqrt{\Delta \tau})$.

By applying each one-body propagator to a Slater determinant
wave function, we will generate another Slater
determinant.\ccite{PhysRevB.41.11352} The many-dimensional
integral/sum over the auxiliary Ising-fields (for each component of
$\vx$ and at each imaginary-time iteration $n$) is performed by Monte
Carlo sampling the fields. The resulting linear combination of Slater
determinants at each iteration $n$ gives a stochastic
representation of the wave function $|\Psi^{(n)}\rangle$. After
convergence, all the sampled determinants from $n\ge n_{\rm eq}$ can
be used collectively to represent the ground state wave function.  The
determinants have to be stabilized periodically with, for example, a
modified Gram-Schmidt orthogonalization
procedure.\ccite{PhysRevB.40.506}

CPMC uses 
importance sampling to steer the sampling toward more ``likely''
auxiliary-field configurations.  This is achieved by a similarity
transformation with a trial wave function $\langle\psi_{T}|\phi\rangle$
to modify the probability density function $p(\vx)$ in
Eq.~(\ref{eq:HS}). The importance sampling does not affect the average
values of the computed observables, only the variance.  The better the
trial wave function $|\psi_{T}\rangle$, the smaller the statistical
error for a fixed amount of Monte Carlo samples.
Since the CPMC process is a branching random walk, population control
has to be applied periodically.\ccite{PhysRevB.55.7464,PhysRevB.57.11446}  As a
manifestation of the variance reduction, the extent of weight
fluctuation (branching) will be reduced with a better
$|\psi_{T}\rangle$.
The use of symmetry-projected trial wave functions leads to a reduction
of the statistical variance, as will be illustrated below.

The sign problem\ccite{SZ_2013LectNotes} is the leading difficulty in
QMC simulations of many-fermion systems.
The problem occurs because
the projection is in general symmetric about $|\phi\rangle$ and
$-|\phi\rangle$. In the random walks, there is no mechanism to
distinguish random walkers of opposite overall signs. In the course of
the projection, if we switched the sign of each random walker (e.g.,by
exchanging two orbitals in $|\phi_\uparrow\rangle$, or
$|\phi_\downarrow\rangle$, in an UHF type simulation, or two
spin-orbitals in a GHF-type simulation), there would be no noticeable
change and the sampled population would give an overlap of opposite
sign with any trial wave function. In the half-filling case of the Hamiltonian in
Eq.~(\ref{eq:HubHamil}), it turns out that symmetry in the propagator
and in $|\Psi_T\rangle$ can make the sign of the overlap remain
non-negative, which eliminates the sign problem.  In general, however,
the overlap with any trial wave function (or with $|\Psi_0\rangle$)
will be zero when averaged over imaginary time (or over the entire
population at any large $n$), leading to infinite Monte Carlo
variance.
The sign problem can be controlled, exactly, by eliminating
the walkers whose overlap with the ground state becomes zero\ccite{SZ_2013LectNotes}
\begin{equation}
 \langle \Psi_{0}|\phi\rangle=0,
\end{equation}
since they will contribute zero at any future projections.  In
practice, in place of the exact ground state wave function, a trial
wave function $|\psi_{T}\rangle$ is used instead to implement the
constraint above,
\begin{equation}
 \langle \Psi_{T}|\phi\rangle=0,
 \label{eq:CP}
\end{equation}
which will introduce a systematic bias in the results.  The importance sampling
transformation imposes this condition naturally, terminating random
walk paths that lead to a zero-overlap with $ |\Psi_{T}\rangle$.  As
mentioned, the systematic error of the constraint tends to be very
small even with simple mean-field trial wave functions.  We will show
below that, with the symmetry-projected wave function, the
constrained-path bias is further reduced and becomes negligible in most of the
systems studied.

\section{Results}
\subsection{Half-filling}
\label{ssec:half-filling}

We first discuss the use of symmetry-projected trial wave functions for
the half-filled Hubbard model.
Since there is no sign problem in this case, CPMC calculations can be
made exact by redefining the important sampling to have a nonzero
minimum.\ccite{PhysRevB.55.7464,PhysRevB.78.165101} If we ignore this
and apply the usual importance sampling with $|\Psi_T\rangle$
literally, the random walks can be constrained to a part of Slater
determinant space because the paths are terminated by the condition in
Eq.~(\ref{eq:CP}). The CPMC results will display a constraint bias. In
this section, we will use the half-filled case in the artificial way
with the constraint as a benchmark system to study the effect of the
symmetry projected trial wave function. Because the exact result is
accessible, this study allows us to systematically examine the effect
of the different symmetries in $|\Psi_T\rangle$.

\begin{figure}[htbp]
  \includegraphics[scale=0.7]{\PLOTFILE{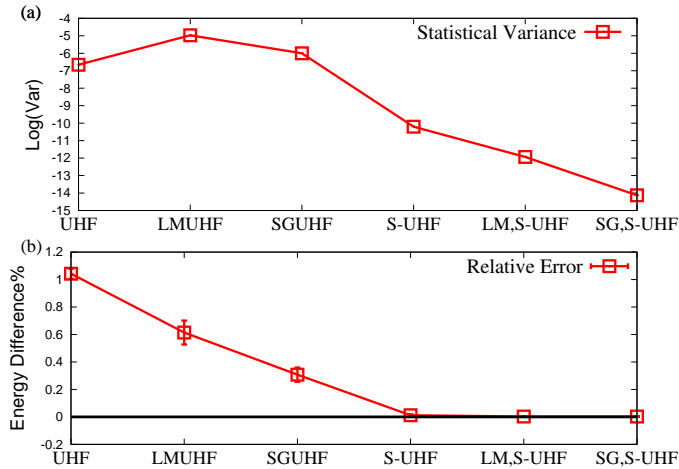}}
  \caption{\label{fig:half-fill} (Color online) Performance of
    different symmetry-projected trial wave functions in CPMC.  Panel
    (a) plots the statistical uncertainty of the computed ground-state energy 
    in a semi-log plot (arbitrary scale).  Panel (b) plots the percentage error,
    $(E_{\rm CP}-E_0)/|E_0|$, between the CPMC energy and the exact
    energy.
    The system is the $4\times4$ Hubbard model, with
    $N_{\uparrow}=N_{\downarrow}=8$ and $U=4$.  The number of walkers
 was 1000, an equilibrium phase of $\beta_{\rm eq}=4$ was
    discarded, and $\Delta\tau=0.01$ was used in the runs. }
\end{figure}

Figure~\ref{fig:half-fill} is an illustration in the $4\times4$
Hubbard model at half-filling ($N_{\uparrow}=N_{\downarrow}=8$) with
$U=4$. Here, the ground state corresponds to a spin singlet with
momentum $(0,0)$ and even parity under all reflections. As we see from the
bottom panel, CPMC with a UHF trial wave function has a bias in the
ground-state energy, with the calculated energy higher than the exact
result ($-13.62185$) by about $\sim 1\%$. As more symmetries are included in the projected UHF based
wave function, the bias is seen to decrease, as well as the
corresponding Monte Carlo variance, shown in the top panel.  The only
exception is in the variance from the UHF trial wave function, which
falls slightly smaller than those from LMUHF and SGUHF trial
wave functions.  The reduction of energy bias and energy variance
indicates that the quality of the trial wave function is
improving. This is born out by their variational energies, which
decrease monotonically following the sequence from left to right (not
shown).

We also see from Fig.~\ref{fig:half-fill} that the CPMC energy
becomes indistinguishable with the exact one as long as the spin
symmetry is imposed.
This is consistent with the findings from our previous studies of
symmetry-preserving trial wave functions which were obtained from a CAS
approach by diagonalizing a truncated active
space.\ccite{PhysRevB.88.125132}

\begin{figure}[htbp]
  \includegraphics[scale=0.03]{\PLOTFILE{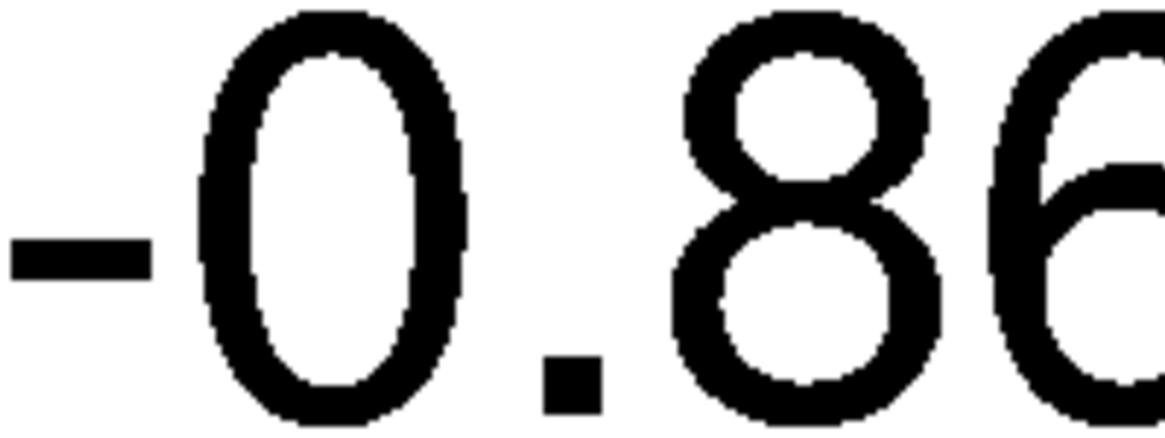}} 
  \caption{\label{fig:half-filling-large} (Color online)
    Size-dependence of the computed energies
  in the half-filled Hubbard model at $U=4$. The behaviors of two
  different forms of trial wave functions are compared. The variational
  energies of the trial wave functions are shown in the upper curves.
  The dashed lines are polynomial fits to the data. CPMC results using
  the two trial wave functions as (artificial) constraints are shown in
  the lower curves, together with \emph{exact} free-projection
  results.  (Note the different vertical scales between the upper and
  lower curves.)  Our best estimate of the ground-state energy in the
  thermodynamic limit is given by the horizontal lines. Its
  statistical uncertainty is indicated by the width of the line in the
  inset.  }
\end{figure}

We next study the behavior of the simulation as a function of system
size.  The results are summarized in
Fig.~\ref{fig:half-filling-large}, which shows the calculated energy per
site from various approaches versus the inverse linear dimension of
the square lattice.  Two types of trial wave functions are compared,
the standard UHF and the spin-symmetry (onto a singlet) projected
S-UHF.  As mentioned earlier, symmetry-projected HF wave functions are
not size-extensive.  The variational energy (per site) of S-UHF is
considerably lower than that of UHF for smaller lattice sizes, but
gradually approaches the UHF value as the system size is increased. As
the fits indicate, the two give the same result in the thermodynamic
limit, $E_{\rm UHF}/L\sim-0.797$.

The CPMC results with an artificial constraint using UHF are shown as
CPMC/UHF. Exact results from free-projection ({\em i.e.}, by
``lifting'' the constraint)\ccite{PhysRevB.88.125132} shown as FPMC/UHF are given for
systems up to $16 \times 16$.  The CPMC/UHF bias is visible at all
lattice sizes.  The largest discrepancy is seen between the two at the
smallest lattice size (rightmost point), corresponding to a $1\%$
error in Fig.~\ref{fig:half-fill}.  Convergence of the ground-state
energy with system size is not monotonic under periodic boundary
conditions, as seen in the bottom set of curves and the inset. To accurately determine the exact
ground-state energy in the thermodynamic limit, we have done
calculations on up to $14\times 14$ lattices with twist-average
boundary conditions, which gives much smaller finite-size effects than
PBC. A polynomial fit\ccite{HusePhysRevB.37.2380} with a leading term of $1/L^{3/2}$ was performed
on the data (not shown), and our best estimate of the energy per site for
the half-filled Hubbard model at $U=4$
as $L\rightarrow \infty$ is $E_0/L=-0.8600(1)$. 
The CPMC/UHF results are seen to converge to a value slightly higher than the exact energy in the thermodynamic limit. 

CPMC using the symmetry-projected S-UHF trial wave function is seen to
agree with exact free projection at all finite lattice sizes studied.  This
extends the conclusion from Fig.~\ref{fig:half-fill} that trial
wave functions with spin symmetry have negligible bias to much larger
supercells.  These data indicate that, despite the lack of
size-extensivity in the S-UHF trial wave function, CPMC seems to
restore a consistent behavior as a function of system size.

\subsection{Away from half-filling}
\label{ssec:away-half-filling}

We next move away from half-filling to examine the effect of the
symmetry-projected HF trial wave function on the sign problem.
We study the system close to half-filling, the region which has the
most severe sign problem. Intermediate values of the interaction
strength $U$ are chosen, comparable to the band width, where the model
is potentially most relevant to strongly correlated materials such as
high-$T_c$
superconductors.

\begin{figure}[htbp]
  \includegraphics[scale=0.7]{\PLOTFILE{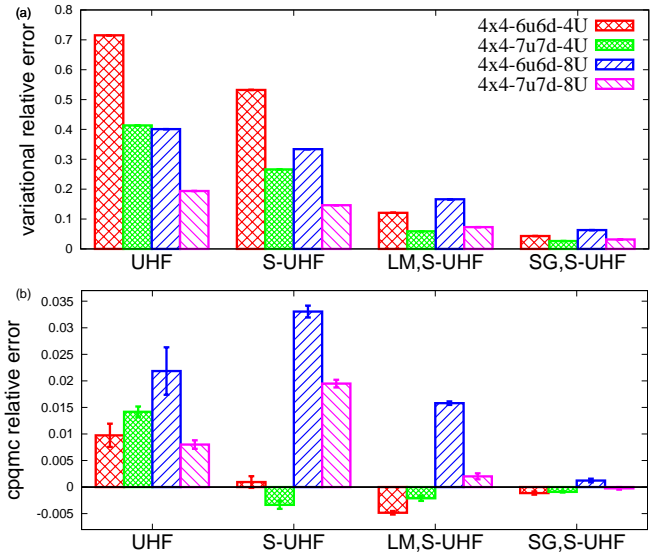}}
  \caption{\label{fig:away-half-fill} (Color online) Variational
    energies (upper panel) of the UHF and three UHF-based
    symmetry-projected wave functions, and the corresponding CPMC
    energies (lower panel) using these as trial wave functions. The
    energies are shown in terms of relative errors in the correlation
    energy (see text).  Note the different scales between the two
    panels.  Four systems are studied, each represented by a different
    color/pattern of vertical bar.  The CPMC results were obtained
    with a population size of $1,000$ and 10 independent blocks of
    size $\beta\sim 1$ each.  The statistical error bars are
    indicated. Trotter error is negligible.
}
\end{figure}

Figure~\ref{fig:away-half-fill} shows the results in $4\times 4$
lattices where exact diagonalizations  are available for
comparison. Two filling values, $N_\uparrow=N_\downarrow=6$ and $7$
are each studied at $U=4$ and $8$. The ground state in all cases is a
spin singlet with momentum $(0,0)$ transforming as the $B_1$ irrep of
the $C_{4v}$ group, save for the 12-electron system at $U=8$, where
the ground state switches to the $A_1$ irrep.  The variational
energies for the UHF wave functions and three symmetry-projected HF
wave functions are shown in the top panel, while the corresponding CPMC
energies are shown in the bottom panel.  Here the energies are shown
in terms of relative errors in the \emph{correlation energy}, defined
as $(E-E_0)/(E_{\rm RHF}-E_0)$, where $E$ is the calculated energy
(either variational or CPMC), $E_0$ is the exact result, and $E_{\rm
  RHF}$ is the restricted HF (equivalent to the Fermi gas
wave function) energy.
The variational energy improves as more symmetries are broken and
restored in the trial wave function, as expected.  The corresponding
CPMC using UHF trial wave functions yields relative errors in the
computed correlation energy of a few percent, while CPMC with
symmetry-projected trial wave functions in general has significantly
smaller errors. CPMC/SG,S-UHF is the most accurate for the four systems
studied here, with relative errors of about $ 0.1\%$ in the
correlation energy. In terms of percentage errors of the total energy
(as plotted in Fig.~\ref{fig:half-fill}), the $U=4$ systems have
errors of $\sim - 0.02\%$
and the maximum error is $0.07\%$ at $U=8$.
Note, that the ground-state energy in CPMC is calculated with the so-called
mixed estimate, which is not variational.\ccite{PhysRevB.59.12788}

\begin{figure}[htbp]
  \includegraphics[scale=0.7]{\PLOTFILE{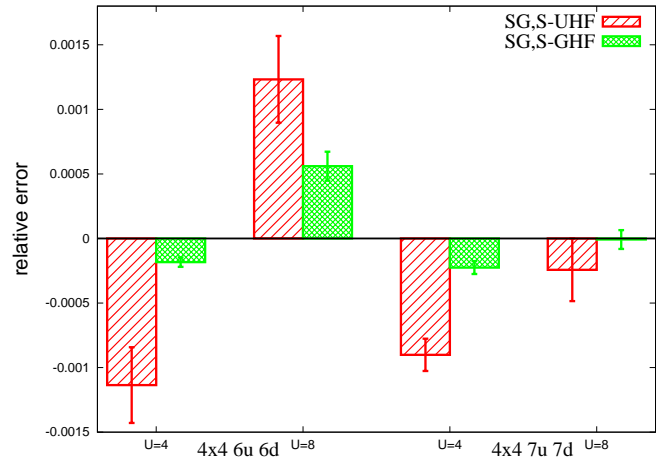}}
  \caption{\label{fig:away-half-fill-ghf} (Color online) Illustration
    of GHF-based symmetry-projected trial wave functions in CPMC.  The
    relative errors in the computed correlation energy from CPMC using
    GHF-type wave functions (SG,S-GHF) are compared with the
    corresponding UHF-type (SG,S-UHF). The latter are taken from the
    rightmost group in Fig.~\ref{fig:away-half-fill}. The same
    computational parameters are used for the two types of calculations.  }
\end{figure}

We have also investigated the effect of GHF-based symmetry-projected
wave functions.
In the usual CPMC calculations discussed thus far, the random walkers
have a form $|\phi_\uparrow\rangle\otimes |\phi_\downarrow\rangle$,
where the $\uparrow$- and $\downarrow$-spin determinants have
$N_\uparrow$ and $N_\downarrow$ orbitals, each specified by $L$
amplitudes, as in UHF.  The one-body propagators in Eq.~(\ref{eq:Bx})
decouple into $\uparrow$- and $\downarrow$-spin components which
operate on $|\phi_\uparrow\rangle$ and $|\phi_\downarrow\rangle$,
respectively.  In the GHF-type, a walker $|\phi\rangle$ contains
$(N_\uparrow + N_\downarrow)$ spin-orbitals, each of which is
specified by $2L$ amplitudes that evolve stochastically.  The one-body
propagators are given by $2N\times 2N$ matrices.  In the present
study, the form of the initial/trial wave function dictates which of
the two approaches is used.

Since there is more freedom in wave functions of symmetry-projected GHF
form,\ccite{PhysRevB.87.235129} lower variational energy can be
obtained. Note that lower energies may only be obtained when the
symmetry-projected states are optimized in the presence of the
projection operators (a VAP approach). This is seen from
Fig.~\ref{fig:away-half-fill-ghf}, which compares CPMC results from
GHF-based and UHF-based trial wave functions with the same
symmetry-projections.  A significant further reduction is seen in the
CPMC results with GHF-based trial wave functions, leading to
essentially exact energies.  In Table~\ref{tab:4x4lattice}, we
summarize all the CPMC ground-state energies with both types of trial
wave functions. 

\begin{table*}
\caption{\label{tab:4x4lattice} CPMC ground-state energies with UHF
  and various symmetry-projected trial wave functions.  Two fillings
  (first column) and $U$ values are studied on a $4\times4$ lattice
  Hubbard model.
The exact results in the last column are from exact diagonalization.
 The statistical error bars in the CPMC results
are on the last digit and shown in parentheses.  }
\begin{ruledtabular}
\begin{tabular}{ccccccccccc}
System &$U$ & UHF &S-UHF& LM,S-UHF&
SG,S-UHF&SG,S$_z$-GHF&SG,S-GHF&exact\\
\hline
$6{\uparrow}~6{\downarrow}$& $4$&-17.703(6)  &-17.727(3) &-17.7428(9) &-17.7327(8) &-17.7316(4)&-17.7301(1) &-17.7296\\
$6{\uparrow}~6{\downarrow}$&$8$&-14.73(4)   &-14.63(1)  &-14.784(3)  &-14.914(3)  &-14.907(2) &-14.920(1)  &-14.925\\
$7{\uparrow}~7{\downarrow}$&$4$&-15.688(4)  &-15.758(3) &-15.753(2)  &-15.7482(5) &-15.7500(8)&-15.7455(2) &-15.7446\\
$7{\uparrow}~7{\downarrow}$&$8$&-11.77(1)   &-11.628(9) &-11.844(7)  &-11.872(3)  &-11.847(3) &-11.8689(9) &-11.8688
\end{tabular}
\end{ruledtabular}
\end{table*}

\begin{figure}[htbp]
  \includegraphics[scale=0.7]{\PLOTFILE{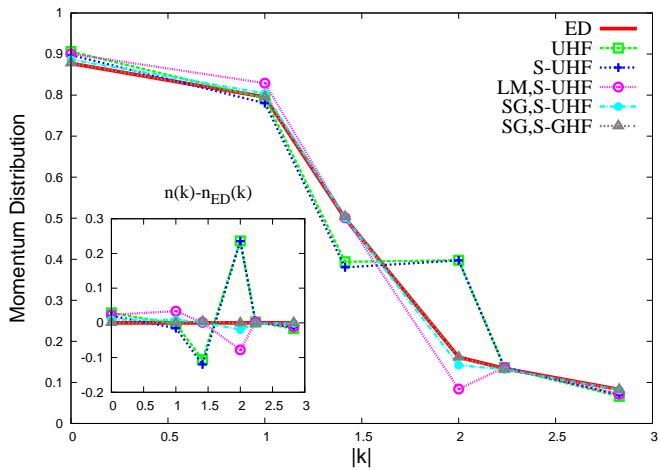}}
  \caption{\label{fig:nk} 
  (Color online) Momentum distribution $n({\mathbf k})$ versus
    $|{\mathbf k}|$. Results from CPMC using different trial
    wave functions are compared with each other and with exact
    diagonalization (ED). The inset shows the error from ED results.
    The \emph{mixed estimator} is used instead
    of back-propagation, to magnify the effect of the trial
    wave function. The system is $4\times4$ with
    $7\uparrow,7\downarrow$ and $U=8$. 
The same run parameters are used as in Fig.~\ref{fig:away-half-fill}. 
}
\end{figure}

The improvements in the energy from CPMC using symmetry-projected HF
wave functions as the constraint show that these wave functions give
better descriptions of the ground-state properties than standard
mean-field. One can then expect that such wave functions will also
improve the calculations of other observables. We study this in
Fig.~\ref{fig:nk} with the example of momentum distribution. In CPMC
calculations of the expectation of observables which do not commute
with the Hamiltonian, the back-propagation
technique\ccite{PhysRevB.55.7464,PhysRevE.70.056702} is almost always
used to obtain a ``pure'' estimator $\langle \Psi_T| e^{-\beta_{\rm
    BP} H} \hat O |\Psi_0\rangle/\langle \Psi_T| e^{-\beta_{\rm BP} H}
|\Psi_0\rangle$, as opposed to the mixed estimate (corresponding to
$\beta_{\rm BP}=0$) which is exact for the energy but biased for a
general operator $\hat O$. For reasonable choices of the
back-propagation time $\beta_{\rm BP}$, the technique gives exact
estimators of any $\langle \hat O\rangle$ except for constrained-path
errors. Here, for illustrative purposes, we use the mixed-estimate to
calculate $n({\mathbf k})$, which helps to magnify the effect of
$\langle \Psi_T|$.

As seen from Fig.~\ref{fig:nk}, the calculated momentum distribution
is incorrect around the Fermi energy when the UHF trial wave function
is used.  This is consistent with the fact that the UHF is a very poor
wave function, and back-propagation must be employed for observables
when such trial wave functions are used.  We see that the SUHF trial
wave function leads to little improvement, despite having improved the
CPMC energy. As more symmetries are included in the projected HF
wave functions, the results improve. The largest step occurs when
space-group is added. This seems reasonable given the open-shell
nature of this system.  The result with the full SG,S-GHF trial
wave function is essentially indistinguishable from exact
diagonalization.

In Table~\ref{tab:biglattice} several CPMC results are shown using
symmetry-projected HF trial wave functions for larger systems away from
half-filling.
Accurate energies from constraint release
calculations\ccite{PhysRevB.88.125132} with CASSCF wave function are
available in these systems, which we use for benchmark and are shown
in the last column.  The UHF trial wave functions are very good at these
fillings and all often difficult to surpass,\ccite{PhysRevB.88.125132}
so it is significant that the symmetry-projected trial wave functions
except the minimal set (spin symmetry
only) perform better. In fact the best results are consistent with the
constraint-release results, with much smaller statistical error bars.

\begin{table*}
\caption{\label{tab:biglattice}CPMC ground-state energy of the Hubbard
  model in square lattices. Symmetry-projected HF trial wave functions
  are used; the ground-state in all cases corresponds to a singlet
  transforming as the $B_1$ irrep of the $C_{4v}$ group with momentum
  $(0,0)$. Results using the UHF trial wave function are listed for
  comparison.
The last column is release constraint\ccite{PhysRevB.88.125132}
results (using the CASSCF trial wave functions), which are the best
estimate of the exact ground-state energy.  The statistical error bars
in the QMC results are on the last digit and shown in parentheses.  }
\begin{ruledtabular}
\begin{tabular}{cccccccccccc}
SYSTEM& $U$ &UHF &S-UHF &LM,S-UHF & SG,S-UHF& SG,S$_z$-GHF& exact \\
$6\times6,~12{\uparrow}\,12{\downarrow}$ & $4$&-42.621(6) &-42.605(3) &-42.633(3)&-42.670(2) &-42.679(1) &-42.669(8)\\
$6\times6,~12{\uparrow}\,12{\downarrow}$ & $8$&-36.68(7) &-36.20(2) &-36.94(1)&-37.188(8) &-37.383(4)&-37.41(6)\\
$8\times8,~22{\uparrow}\,22{\downarrow}$ & $4$&-75.886(8) &-75.875(7) &-75.900(6)&-75.904(3)& &-75.893(9)
\end{tabular}
\end{ruledtabular}
\end{table*}

Computational scaling with system size is an important consideration
for any many-fermion method.  In a CPMC calculation, the primary
objects involving the trial wave function that need to be calculated
repeatedly are the overlap $\langle \Psi_T|\phi\rangle$ and the
one-body Green function $\langle \Psi_T|c_{\bf{k}}^\dagger
c_{\bf{l}}|\phi\rangle/\langle \Psi_T|\phi\rangle$.  For a single-determinant
$|\Psi_T\rangle$ such as UHF, the computations\ccite{SZ_2013LectNotes}
scale like regular mean-field calculations.  For a $|\Psi_T\rangle$ in
the form of Eq.~(\ref{SP-PHF}), a straightforward way is to expand it
as a linear combination of Slater determinants after the reference
determinant and variational parameters have been determined.  This is
the approach we have taken in the present paper as a proof-of-concept study of
the effectiveness of such wave functions. However, the number of Slater
determinants in such an expansion grows rapidly with system size.  For
example, in the $8 \times 8$ calculation in
Table~\ref{tab:biglattice}, the SG,S-UHF trial wave function involved
$14,336$ determinants.  An alternative is to use 
Eq.~(\ref{SP-PHF}), and incorporate the symmetry projection
in the QMC, computing the objects with numerical integrations
over the projection operators.\ccite{PhysRevB.85.245130} The
computational cost is then proportional to that for a simple HF trial
wave function, with a proportionality constant which depends on the
symmetries involved but only grows modestly with system size (e.g.,
linearly for translation). This is a very appealing feature of the
symmetry-projected HF wave functions as $|\Psi_T\rangle$, in contrast
with, for instance, the CASSCF-type of wave functions, and offers the
promise of scalable calculations with our approach to reach large
system sizes.

\section{Conclusion}
We have studied symmetry-projected HF wave functions as trial
wave functions for CPMC. The symmetries are restored by projecting from
an UHF or GHF type wave function, minimizing the variational energy
after the projection.  This gives a hierarchy of wave functions with
increasing quality.  The CPMC results using these as trial
wave functions generally become increasingly more accurate, and the
Monte Carlo variance increasingly smaller for the same amount of
samples.  It is shown that CPMC largely restores size-extensivity in
symmetry-projected HF wave functions.  At half-filling in the Hubbard
model, projected wave functions with spin symmetry  provide
an excellent description of the systems.  Away from half-filling, the
projected wave functions with space group symmetry and spin symmetry
lead to highly accurate results, often with the constrained-path
systematic error less than $0.1\%$ in the \emph{correlation energy}.
Wavefunctions projected from a GHF-type reference tend to perform
even better compared with the ones projected from UHF-types.

As discussed above, the symmetry-projected wave function can be
implemented in CPMC as trial wave functions with a computational cost
which scales modestly with system size.  The development in this paper
is thus potentially a scalable and systematically improvable quantum
Monte Carlo approach for extended systems. Moreover, one can further
increase the quality of the symmetry-projected trial wave function by
doing particle number symmetry breaking or by incorporating
multi-component expansions such as those considered in our previous
work.\ccite{PhysRevB.87.235129}
We finally note that the current developments in CPMC have paved the
way for accurate many-body calculations of Hamiltonians with
spin-orbit coupling.
\begin{acknowledgments}
We thank
S.~Chiesa,
F.~Ma,
W.~Purwanto,
and E.~Walter
for helpful discussions.
This research was supported by DOE (Grant No.~DE-SC0008627 and
DE-FG02-11ER16257) and NSF (Grant No.~DMR-1006217). The work at Rice
University was supported by DOE, Office of Basic Energy Sciences,
Heavy Element Chemistry program (Grant No.~DEFG02-04ER15523) and
DOE-CMCSN (Grant No.~DE-SC0006650).  We also acknowledge
an INCITE award for computing using resources of the Oak Ridge
Leadership Computing Facility at the Oak Ridge National Laboratory,
which is supported by the Office of Science of the U.S. Department of
Energy under Contract No.~DE-AC05-00OR22725 and CPD at College of William and Mary. G.E.S. is a Welch
Foundation Chair (C-0036).

\end{acknowledgments}

\bibliography{afqmc}

\end{document}